\documentstyle[11pt,aaspp4,epsfig,psfig]{article}       

\newcommand{\etal}{{\it et al.\/}}
\newcommand{\be}{\begin{equation} }
\newcommand{\ee}{\end{equation} }
\newcommand{\bea}{\begin{eqnarray} }
\newcommand{\eea}{\end{eqnarray} }

\def\dert#1#2{{{d#1}\over{d#2}}}
\newcommand{\kes}{\kappa_{es}}
\newcommand{\kff}{\kappa_{ff}}

\lefthead{Zane, Turolla, \& Treves}
\righthead{MAGNETIZED ATMOSPHERES AROUND ACCRETING NEUTRON STARS}

\begin{document}

\title{Magnetized Atmospheres around Neutron Stars \\ Accreting at
Low Rates}

\author{Silvia Zane}
\affil{Nuclear and Astrophysics Laboratory,
University of Oxford, \\ Keble Road, Oxford OX1 3RH, England\\
e--mail: zane@astro.ox.ac.uk}
\author{Roberto Turolla}
\affil{Dept. of Physics, University of Padova, \\ Via Marzolo 8, 35131 Padova,
Italy \\ e--mail: turolla@pd.infn.it}
\and
\author{Aldo Treves}
\affil{Dept. of Sciences, University of Insubria, \\
Via Lucini 3,
22100, Como, Italy \\ e--mail: treves@uni.mi.astro.it}

\begin{abstract}

We present a detailed investigation of atmospheres around
accreting neutron stars with high
magnetic field ($B\gtrsim 10^{12}$ G) and low luminosity ($L\lesssim
10^{33}$  erg/s). We compute the
atmospheric structure, intensity and emergent spectrum for a
plane--parallel, pure hydrogen medium by solving the transfer equations
for the normal
modes coupled to the hydrostatic and energy balance equations. 
The hard tail found in previous investigations for accreting,
non--magnetic neutron stars with comparable luminosity is suppressed and
the X--ray spectrum, although still harder than a blackbody at the star
effective temperature, is nearly planckian in shape. Spectra from
accreting atmospheres, both with high and low fields, are found to exhibit
a significant excess at optical wavelengths above the Rayleigh--Jeans tail
of the X--ray continuum.

\end{abstract}


\section{Introduction}\label{intro}

The problem of investigating the properties of radiation emitted
by neutron stars (NSs) accreting at low rates, $\dot M\approx
10^{10}-10^{14}$ g/s, became of interest after it was realized
that the Galaxy may contain a large population of low luminosity
magnetic accretors (see e.g. Nelson, Wang, Salpeter, \& Wasserman
\cite{nel95:1995}). The Galaxy should harbor more than $10^3$
Be/X--ray binaries with an accreting NS shining at $\approx
10^{32}-10^{34}$ erg/s (Rappaport, \& van den Heuvel
\cite{rvdh82:1982}; van den Heuvel, \& Rappaport
\cite{vdhr86:1986}). Moreover, assuming a supernova birth rate of
10--100 yr$^{-1}$, $\sim 10^8-10^9$ old, isolated NSs (ONSs)
should be present in the Galaxy. Accretion onto a strongly
magnetized, moving neutron star may be severely hindered for
different reasons, but there is the possibility that a small,
albeit non--negligible, fraction of ONSs may be accreting directly
from the interstellar medium and some of them might be above the
sensitivity threshold of ROSAT (e.g. Treves, \& Colpi
\cite{tc91:1991}; Blaes, \& Madau \cite{bm93:1993}; see Treves,
Turolla, Zane, \& Colpi \cite{review:1999} for a review).

At variance with neutron stars accreting at high rates, e.g. in X--ray
pulsators, in low--luminosity sources the interaction of the escaping
radiation with the inflowing material is of little importance, so they
provide a much simpler case for investigating the physics of accretion in a
strongly magnetized environment. For luminosities far below the Eddington
limit, the accretion problem becomes
germane to that of calculating
the spectrum emerging from static atmospheres around cooling NSs.
Spectra from cooling NSs have been widely investigated by a number of
authors in connection with the X--ray emission from young, millisecond
pulsars and isolated NSs, both for low
and high magnetic fields and for different chemical compositions (see e.g.
Romani \cite{rom87:1987}; Shibanov, Zavlin, Pavlov \& Ventura
\cite{sh92:1992}; Rajagopal, \& Romani \cite{rr96:1996};
Pavlov, Zavlin, Tr\"umper,
\& Neuh\"auser \cite{pztn96:1996}). Emerging
spectra are not very different from a blackbody at the star effective
temperature, the
distinctive hardening present at low fields ($B\lesssim 10^9$ G) becoming
less pronounced when the magnetic field is $\sim
10^{12}-10^{13}$ G. Similar conclusions were reached for the
spectrum emitted by low--luminosity, low--field, accreting NSs by
Zampieri, Turolla, Zane, \& Treves (\cite{ztzt95:1995}, hereafter
ZTZT) for a pure hydrogen atmospheric composition.

The search of isolated neutron stars with ROSAT produced in recent
years half a dozen promising candidates (Walter, Wolk, \& Ne\"uhauser
\cite{wwn96:1996}; Haberl \etal \ \cite{ha97:1997}; Haberl, Motch \&
Pietsch \cite{hmp98:1998};
Schwope \etal \ \cite{sch99:1999}; Motch \etal \ \cite{mo99:1999};
Haberl, Pietsch \& Motch \cite{hpm99:1999}). All
of
them show a soft, thermal X--ray spectrum, with typical energies
$\sim 100$ eV, and have an exceedingly large X--ray to optical flux ratio,
$\gtrsim 10^4$. Although their association with isolated neutron stars is
firmly established, their interpretation in terms of an accreting or
a cooling object is still a matter of lively debate.
Present models predict rather similar spectral distributions
in both cases, especially in the X--ray band. It is therefore of
particular importance
to improve our theoretical understanding of these two classes of sources,
looking, in particular, for spectral signatures which can enable us to
discriminate between them.

In this paper we present a first detailed calculation of spectra
emitted by strongly magnetized ($B\gtrsim 10^{12}$ G), accreting neutron
stars, focusing our attention
on low luminosities, $L \sim 10^{30}$--$10^{33}$ erg/s, such as those expected
from old neutrons stars accreting the interstellar medium.
Spectral distributions are computed
solving the transfer equations for the normal modes coupled to the
hydrostatic equilibrium and
the energy balance for different values of the accretion
rate and the magnetic field.

We find that spectra emerging from magnetized, accretion
atmospheres are blackbody--like in the X--ray band, in close
agreement with the known results for cooling, magnetized
atmospheres. However, accretion spectra show a new and distinctive
feature at low energies, being characterized by an excess over the
Raleigh--Jeans tail of the X--ray continuum below $\sim 10$ eV.
The same behaviour is found in accretion atmospheres around
unmagnetized neutron stars, but, as already pointed out by ZTZT,
the X--ray spectrum is sensibly harder in this case. This result
may be relevant in connection with the isolated neutron star
candidate RX J18563.5-3754. Multiwavelength observations of this
source indicate that, while ROSAT data are well fitted by a
blackbody at $T_{eff}\sim 60$ eV, HST points lie above the
extrapolation of the fit in the optical (Walter, \& Matthews
\cite{wm97:1997}).

The plan of the paper is as follows. The input physics relevant to
our model is presented in \S \ref{model}: radiative transfer in a
magnetized plasma is discussed in \S \ref{model-radtra} and the
structure of an accreting, magnetized atmosphere in \S
\ref{model-struct}. Computed spectra are presented in \S
\ref{results}. Discussion and conclusions follow in \S
\ref{discuss}.

\section{The Model}\label{model}

\subsection{Radiation Transfer}\label{model-radtra}

In this paper we consider a magnetized, nondegenerate, pure
hydrogen, cold plasma, in which the main radiative processes are
free--free emission/absorption and Thomson scattering. The plasma
is in local thermal equilibrium (LTE) at temperature $T$. We
consider a plane--parallel geometry with normal ${\bf n}$ parallel
to the magnetic field ${\bf B}$ and to the $z$--direction. The
stratification of the atmosphere is described by using as a
parameter the scattering depth $\tau$, as defined for an
unmagnetized medium \be \tau = \kes \int_z^{\infty} \rho\, dz \ee
where $z$ is the coordinate variable, $\rho$ is the plasma
density, $\kes = \sigma_T/m_p$ is the Thomson opacity and
$\sigma_T$ is the Thomson cross section.

In the following, we neglect collective plasma effects and consider only
the limit $\omega_p^2/\omega^2 \ll 1$, where $\omega_p = (4 \pi n_e e^2
/m_e)^{1/2}$
is the plasma frequency and $n_e$ electron density.
We also consider only frequencies lower than the
electron
cyclotron frequency, $\omega_{c,e} = eB/m_ec$, so the semitransverse
approximation can be assumed to hold. Since, for $\tau \gtrsim 0.01$, the
temperature in the
atmosphere is always
$\lesssim 10^7$ K (see \S \ref{model-struct} and ZTZT) and
scattering dominates over true absorption only for
$\tau\sim 1$, Comptonization
is negligible. For this reason, similarly to what is done for cooling,
magnetized
atmospheres (see e.g. Shibanov, \etal \ \cite{sh92:1992}), only
conservative scattering is
accounted for in the transfer equations (see, however, the discussion in
\S \ref{model-struct} for the role of Compton heating/cooling in
the energy balance
of the external atmospheric layers).

Under these assumptions, the coupled equations for the transfer of the
two normal modes take the form (see e.g. Gnedin \& Pavlov
\cite{gp74:1974}; Yahel \cite{y80:1980})
\bea
\label{tr1}
- y_G \mu \dert{f^1}{\tau} & = &
\int K_s^{11} f^{1'} d \mu'
+ \int K_s^{21}  f^{2'} d \mu' -
 k_s^{1}  f^1
+ k_{ab}^1  \left ( { 1 \over 2} b_\nu - f^1 \right )\nonumber\\
- y_G \mu \dert{f^2}{\tau} &=&
\int K_s^{22}  f^{2'} d \mu'
+ \int K_s^{12}  f^{1'} d \mu' -
 k_s^{2}  f^2
+ k_{ab}^2  \left ( { 1 \over 2} b_\nu - f^2 \right )
\eea
where $y_G = \sqrt{1 - 2GM/Rc^2}$ is the gravitational redshift
factor, $R$ and $M$ are the star mass and radius,
$\mu = {\bf n}\cdot{\bf s}=\cos \theta$, $f^i \left ( \tau,
\mu, \nu \right )= c^2 I^i/2h^4\nu^3$ denotes the photon occupation number
for the ordinary ($i =1$) and extraordinary ($i=2$) mode, $I^i$ is the
specific intensity, $b_\nu =
c^2B_\nu/2h^4\nu^3 $, and $B_\nu$
is the Planck function.
In equations (\ref{tr1}) $k_{ab}^i$ is the total free--free opacity,
\be
k_s^i(\mu,\nu) = \sum_{j=1}^2 \int K_s^{ij}\, d\mu'
\ee
and $K_s^{ij}(\mu,\mu',\nu)$ is the probability that an incident photon, which
has polarization $\hat e^i$ and propagates in the direction $\mu$,
scatters into a direction $\mu'$ and polarization $\hat e^j$.
All the quantities appearing in equations (\ref{tr1}) are referred to the
local observer, at rest on the star surface; the photon energy measured by
an observer at infinity is given by $h\nu_\infty = y_Gh\nu$.
The integral
terms appearing into equations (\ref{tr1}) account for the scattering
emissivities, and
all opacity/emissivity coefficients are normalized to $\kes$.
The expressions for the opacities
relevant to the present calculation are reported in appendix
\ref{appa}.

\subsection{Atmospheric Structure}\label{model-struct}

Accretion atmosphere models are constructed by solving the transfer
equations (\ref{tr1}) coupled to the hydrostatic equilibrium and
the energy equation.
The hydrostatic balance is simply expressed as
\be
\label{pre}
\dert{P} {\tau } = {GM \over y_G^2 R^2 \kes } \, ,
\ee
where $P =  k \rho T/\mu_em_p$ ($\mu_e\sim 1/2$ for completely
 ionized hydrogen)
 is the gas pressure and we consider only the case $L/L_{Edd} \ll 1$,
where
 $L=L(\tau)$ is the luminosity measured by the local observer. Since in
 all our models the ram pressure of the accreting material, $1/2 \rho
v^2$,
 turns out to be much smaller than the thermal pressure, it has
 been neglected, together with the radiative force. In this    
 limit equation (\ref{pre}) is
 immediately integrated, and gives the density as a function of depth
 \be\label{dens}
 \rho = \frac{GMm_p}{2 y_G^2 R^2\kes}\frac{\tau}{kT(\tau)}\,.
 \ee

The energy balance just states that the net radiative cooling must
equate the heating $W_H$ supplied by accretion. The radiative
energy exchange is obtained adding equations (\ref{tr1}) together,
after multiplying them by $(\hbar\omega)^3$, and integrating over
angles and energies. Since we assumed conservative scattering, its
contribution to the energy balance clearly vanishes. However, as
discussed in previous investigations (Alme, \& Wilson
\cite{aw73:1973}; ZTZT), Comptonization is ineffective in
modifying the spectrum, but plays a crucial role in determining
the temperature in the external, optically thin layers. Contrary
to what happens in cooling atmospheres, the temperature profile
shows a sudden rise (or ``jump'') in the external, low--density
layers where the heating produced by the incoming protons is
mainly balanced by Compton cooling. Including Compton
heating/cooling the energy equation becomes
\be
\label{en2}
k_P  \left ( \frac{aT^4}{2} - {k_{am}^1 \over k_P } U^1
- {k_{am}^2 \over k_P } U^2 \right ) + \left
(\Gamma - \Lambda \right)_C
= {W_H \over c \kes}
\ee
where $U^i$ is the radiation energy density of mode $i$ and $k_P$,
$k^i_{am}$ are defined in strict analogy with the Planck and absorption
mean opacities in the unmagnetized case.
In evaluating the previous expression the approximated formula by Arons,
Klein, \& Lea (\cite{akl87:1987}) for the Compton rate in a magnetized
plasma, $(\Gamma - \Lambda)_C$, has been used.

The detailed expressions for the heating rate $W_H$ and the stopping depth
$\tau_B$ in a magnetized atmosphere are presented in appendix \ref{appb}.

\section{Results}\label{results}

\subsection{Numerical Method}\label{numeth}

The numerical calculation was performed adapting
to radiative transfer in a magnetized medium the tangent--ray
code developed by Zane,
Turolla, Nobili, \& Erna (\cite{crm:1996}) for
one--dimensional, general--relativistic radiation transfer. The method
performs an ordinary $\Lambda$--iteration for computing the scattering
integrals.
Schematically, the calculation proceeds as follows.
First an initial temperature profile is specified (usually that of an
unmagnetized model with similar parameters calculated by ZTZT) and the
zero--th order approximation for $f^1$, $f^2$ is computed solving
equations (\ref{tr1}) with no scattering emissivity.
The boundary conditions for ingoing ($\mu < 0$) trajectories are
$f^i = 0$ at $\tau = \tau_{min}$
while diffusive boundary conditions at $\tau = \tau_{max}$ were used for
outgoing ($\mu > 0$) trajectories.
The computed intensities are then used to evaluate the scattering
integrals and the whole procedure is repeated,
keeping the temperature profile unchanged. As soon as corrections on
the intensities are small enough, new temperature and density profiles
are obtained solving equations (\ref{en2}) and (\ref{dens}). The whole
scheme is then iterated to convergence. Each model is
completely characterized by the
magnetic field strength, the total luminosity and the luminosity at
$\tau_B$, or, equivalently, by $B$ and the heating rate $W_H$ (see
appendix \ref{appb}). Since the code solves the full transfer
problem, it allows for the complete determination of the
radiation field, including its angular dependence. This ensures a more
accurate treatment of
non--anisotropic radiative process with respect to angle--averaged,
diffusion approximations. Owing to
the gravitational redshift, the total accretion luminosity measured at
infinity is related to the local luminosity at the top of the
atmosphere by $L_\infty = y_G^2 L(0)$.

Models presented below were computed using a logarithmic grid with
300 equally--spaced depth points, 20 equally--spaced angular points and
48 energies. We explored a wide
range of luminosities and considered
two representative values of the magnetic field, $B= 10^{12}$ and
$10^{13}$ G. The model parameters are reported in Table 1, together with
the accretion rate
\be
\dot M = L(0) { \left [ 1 - L(\tau_B)/L(0) \right ] \over \eta
c^2}\,;
\ee
here $\eta= 1 - y_G$ is the relativistic efficiency.
Since
the values of $\tau_B$ and $\omega_{c,e,p}$ depend on $B$, the adopted
boundaries in depth and energy
vary from model to model.
Typical values are $\tau_{max} \sim 10^2$ and $\tau_{min}
\sim 10^{-6}-10^{-8}$; the energy range goes from 0.16 eV to 5.45 keV.
The angle--averaged effective depth is always $\gtrsim 100$ at
$\tau_{max}$ and $\sim 10^{-5}$ at $\tau_{min}$.

Convergence was generally achieved after 20--30 iterations with a
fractional accuracy
$\sim 0.01$ both in the hydrodynamical variables and in the radiation
field.
As a further check on the accuracy of our solutions,
the luminosity evaluated numerically was compared with
\be
\label{lumi}
L(\tau) \approx  L(0) - \left [ L(0)  - L ( \tau_B ) \right ]
{{1 - [1 - (1-v_{th}^4/v_{ff}^4)(\tau/\tau_B)]^{1/2}}\over{1 -
v_{th}^4/v_{ff}^4}}
\ee
which follows integrating the first gray moment equation
\be
\dert { L} {\tau} = - { 4\pi R^2 f_A W_H
\over
y_G \kes }
\ee
and provides an analytical expression for $L$ at $\tau <  \tau_B$ (see
appendix \ref{appb} for notation). Within
the range of validity of equation (\ref{lumi}), the two values of $L$ differ
by less
than 4\%.

In all models $M= 1 M_\odot$, $R = 6 GM/c^2\simeq 0.89 \times 10^6$ cm
which correspond to $\tau_s \simeq 3.3$ (see equation [\ref{taus}]). Only
model A5, the unmagnetized one, was computed with $\tau_s \simeq 8$. One
of the largest complication introduced by the presence of the magnetic
field is the large--scale pattern of the accretion flow. In particular,
when the accretion rate is small the way the spherically symmetric
infalling plasma enters the magnetosphere is not fully understood as yet
(see e.g. Blaes, \& Madau \cite{bm93:1993}, Arons \& Lea
\cite{al80:1980}). In order to bracket uncertainties, we assume a fiducial
value for the fraction of the star surface covered by accretion, $f_A =
0.01$. We stress that this choice has no effect on the spectral properties
of the emitted radiation and affects only the total luminosity, which
scales linearly with $f_A$.

\subsection{Emerging Spectra}\label{spectra}

\begin{figure}
\centering
\psfig{file=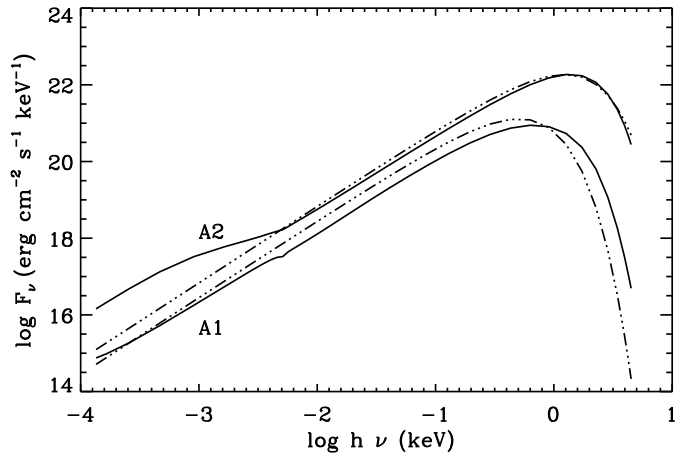,width=10cm}
\caption{Emergent spectra for models A1 and A2 (full lines),
together with the
blackbody spectra at the neutron star effective temperature (dash--dotted
lines).
\label{spe12}}
\end{figure}

\begin{figure}
\centering
\psfig{file=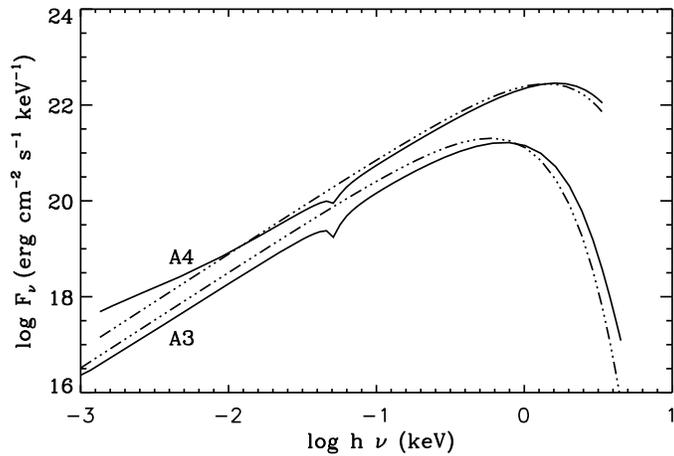,width=10cm}
\caption{Same as in
figure \ref{spe12} for models A3 and A4.
\label{spe13}}
\end{figure}

\begin{figure}
\centering
\psfig{file=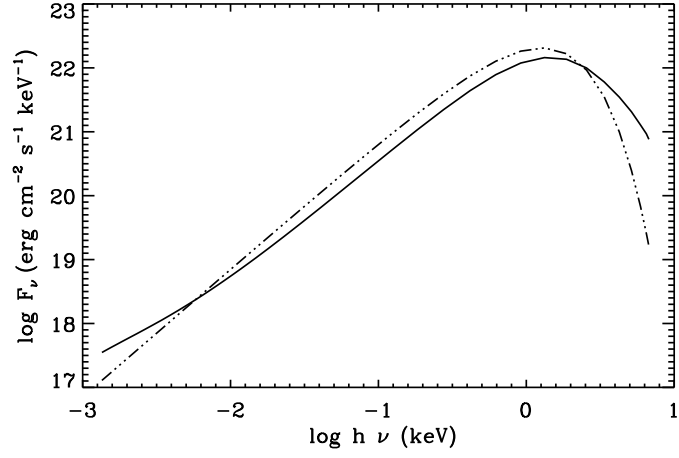,width=10cm}
\caption
{Same as in figure \ref{spe12} for $B = 0$ (model A5).
\label{spe0}}
\end{figure}

\begin{figure}
\centering
\psfig{file=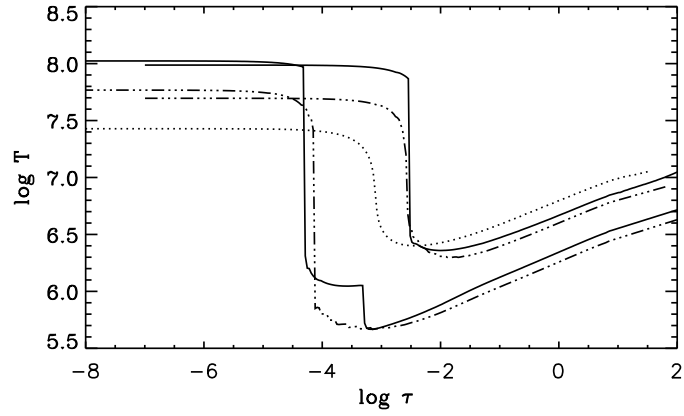,width=10cm}
\caption
{The run of the gas temperature in the atmosphere for the models of
figures \ref{spe12},
\ref{spe13} and \ref{spe0}; solid lines correspond to  $B =
10^{12}$ G, dash--dotted lines to $B = 10^{13}$
G and dotted lines to $B=0$.\label{tempfig}}
\end{figure}

The emergent spectra for models A1--A2 and A3--A4 are
shown in figures \ref{spe12} and \ref{spe13}, together with the blackbody at
the NS effective
temperature, $T_{eff}$. An unmagnetized model (A5) with similar
luminosity is shown in
figure \ref{spe0}. The corresponding temperature profiles are
plotted in figure \ref{tempfig}. Note that in all the plots the photon
energy is already corrected for the gravitational redshift, so the
spectral distribution is shown as a function of the energy as observed
at Earth. As it is apparent comparing the different curves (see also
the solutions  computed by ZTZT)
the thermal stratification of the atmosphere shows the same general
features (inner layers in LTE, outer region dominated by Comptonization)
independently of $B$.

The sudden growth of $T$ (up to $10^7-10^8$ K) that appears in the
external layers is basically due to the fact that
free--free cooling can not balance the heating produced by
accretion at low densities. The temperature then rises until it
reaches a value at which Compton cooling becomes efficient. This
effect has been discussed by ZTZT and Zane, Turolla, \& Treves
(1998) in connection with  unmagnetized atmospheres and they have
shown that the temperature jump is located at the depth where the
free--free  and Compton thermal timescales become comparable,
$t_{ff}\sim t_C$. In the magnetized case the situation is very
similar, but now there are two relevant free--free timescales,
$t^{(1)}_{ff}$ and $t^{(2)}_{ff}$, one for each mode. Comptonization
becomes the  dominant cooling process for $
t_C\lesssim \min(t^{(1)}_{ff}, t^{(2)}_{ff})$
and gives rise to
the large jump present in all accretion models. For some values of the
model parameters, the
region where $ \rho = \rho_{vac}$ coincides with the photospheric
region for both modes (see appendix \ref{vacuum}). In this case,
vacuum effects can produce the
peculiar ``double jump'' structure present in model A1, with the
first (small) jump located at a depth where $t^{(1)}_{ff}/
t^{(2)}_{ff}\sim 1$.

In order to compare our results with models available in the
literature, some cooling atmospheres have been also computed
setting $W_H = 0$ in equation (\ref{en2}); accordingly, the
luminosity is now a constant. The emergent spectra for two such
models (C1 and C2, see Table 1) show a good agreement with those
computed by Shibanov \etal \ (\cite{sh92:1992}) using the
diffusion approximation.

We find that the spectral hardening at low luminosities, typical of
unmagnetized atmospheres, is far less pronounced but still present up to
field strengths $\sim 10^{13}$ G and tends to
disappear at large enough luminosities. For comparison, with $L
\sim 4 \times 10^{33}$ erg/s, the magnetized ($B=10^{12}$ G) spectrum has
negligible hardening while the hardening ratio is still $\sim 1.6$ in
unmagnetized
models of similar luminosity (ZTZT). The overall dependence of
the continuum is not particularly sensitive to
the value of the magnetic field, although the absorption feature at the
proton cyclotron energy becomes more prominent with increasing $B$.

\subsection{The Optical Excess}\label{excess}

\begin{figure}
\centering
\psfig{file=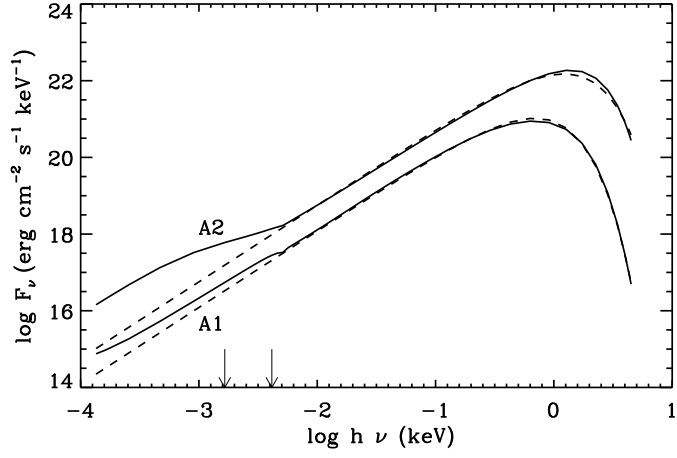,width=10cm}
\caption
{Same models as in figure \ref{spe12},
together with the two blackbody functions which best--fit the X--ray
spectra in the 0.03--4.5 keV interval (dashed lines). The arrows mark
the 3000--7500 A band. \label{spe12fit}}
\end{figure}

\begin{figure}
\centering
\psfig{file=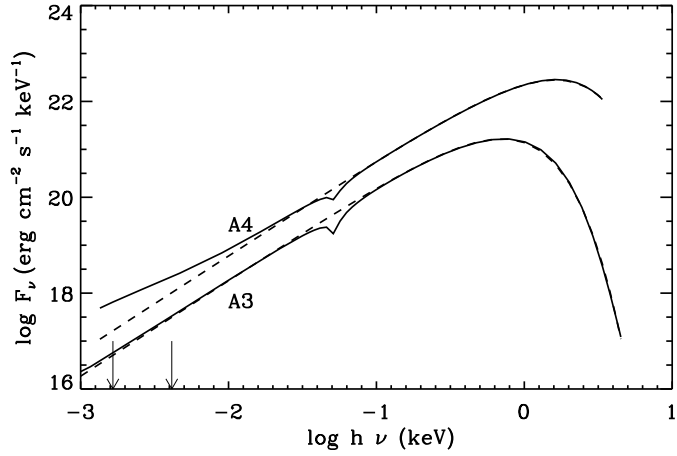,width=10cm}
\caption
{Same as in figure \ref{spe12fit}, for the models with
$B=10^{13}$ G; here the fit is computed in the 0.19--3.4/4.5 keV
band for the two solutions.
\label{spe13fit}}
\end{figure}

\begin{figure}
\centering
\psfig{file=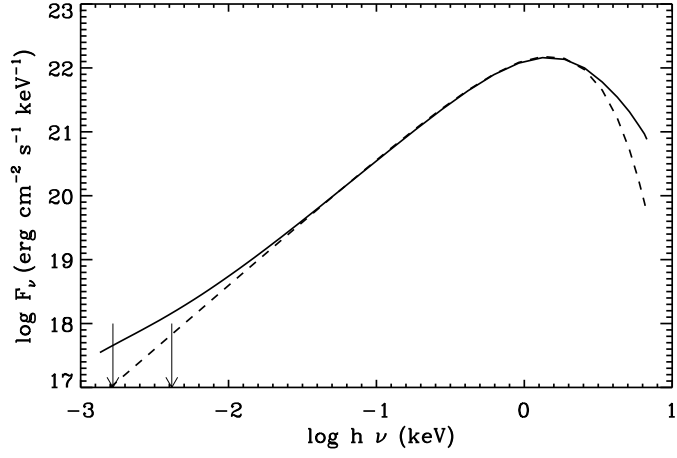,width=10cm}
\caption
{Same as in figure \ref{spe12fit}, for the models with
$B=0$; here the fit is computed in the 0.02--3.3 keV
band. \label{spe0fit}}
\end{figure}

The most striking result emerging from our computations is that,
although spectra from cooling and accreting H atmospheres are
rather similar in the X--rays,  they differ substantially at low
energies. Below $\sim 10$ eV spectra from accreting atmospheres
exhibit a soft excess with respect to the blackbody spectrum which
is not shared by the cooling models. This feature, which is
present also in the unmagnetized case, can be viewed as a
distinctive spectral signature of a low--luminosity, accreting
neutron star. The fact that it was not reported by ZTZT (and by
previous investigators, see e.g. Alme, \& Wilson \cite{aw73:1973})
is because they were mainly interested in the shape of the X--ray
continuum and their energy range was not large enough to cover the
optical band; besides, some numerical problems, related to the
moment formalism used to solve the transfer, prevented ZTZT to
reach very low frequencies. The evidence of the soft excess is
even more apparent in figures \ref{spe12fit}, \ref{spe13fit} and
\ref{spe0fit} where synthetic spectra are plotted together with
the best fitting blackbody in the X--ray band. The excess at two
selected optical wavelengths ($\lambda = 3000\,, 6060$ A, see
discussion below), together with the temperature $T_{fit}$ of the
best--fitting blackbody in the X--ray band, are reported in Table
2.

The appearance of an optical excess in accreting
models is related to the behaviour of the
temperature, which is different from that of cooling models.
The external layers are now
hotter because Comptonization dominates the thermal balance there.
The low energy tail of the spectrum decouples at a depth that corresponds
to the (relatively) high temperatures near the jump, and emerges at
infinity as a planckian at a temperature higher than $T_{eff}$. By
decreasing the luminosity, the temperature jump moves at lower
scattering depth and the frequency
below which the spectrum exceeds the blackbody at $T_{eff}$ becomes
lower.

The presence of an optical excess has been reported in the
spectrum of a few isolated, nearby pulsars (Pavlov, Stringfellow,
\& C\'ordova \cite{psc96:1996}) and, at lower luminosities, in the
spectrum of the ONSs candidate RX J18563.5-3754. RX J18563.5-3754
was observed in the X--ray band with ROSAT (Walter, Wolk, \&
Neuh\"auser \cite{wwn96:1996}) and by HST at $\lambda = 3000$  and
6060 A (Walter, \& Matthews \cite{wm97:1997}). These
multiwavelength observations made evident that the spectrum of RX
J18563.5-3754 is more complex than a simple blackbody. The
blackbody fit to PSPC data underpredicts the optical fluxes
$f_{3000}$ and $f_{6060}$ by a factor 2.4 and 3.7 respectively
(Walter, \& Matthews \cite{wm97:1997}; see also Pavlov \etal \
\cite{pztn96:1996}). Models of cooling atmospheres based on
different chemical compositions also underestimate the optical
fluxes (Pavlov \etal \ \cite{psc96:1996}), while models with two
blackbody components or with a surface temperature variation may
fit the $f_{3000}$ flux.
Recent spectra from non--magnetic atmospheres with Fe or Si--ash
compositions (see Walter, \& An \cite{wa98:1998}) may also provide
a fit of both the X--ray and optical data, although these spectral
models agree with those of Rajagopal, \& Romani (\cite{rr96:1996})
but not with those of Pavlov \etal \ (\cite{pztn96:1996}). Given
the considerable latitude of the unknown parameters $f_A$, $L$,
$B$, however, results presented here indicate that the full spectral
energy distribution
may be consistent with the picture of an accreting NS. We want
also to note that all models computed here have a blackbody
temperature higher than that required to fit the X--ray spectrum
of RX~J18563.5-3754 (see Table 2) and, although we are far from
having explored the model parameter space, present results
indicate that the excess decreases for decresing luminosities.
Although the optical identification of another isolated NS
candidate, RX J0720.4-3125, still lacks a definite confirmation,
it is interesting to note that the counterpart proposed by
Kulkarni \& van Kerkwijk (\cite{kvk98:1998}) also shows a similar
excess.

\subsection{Fraction of Polarization}\label{polariz}

\begin{figure}
\centering
\psfig{file=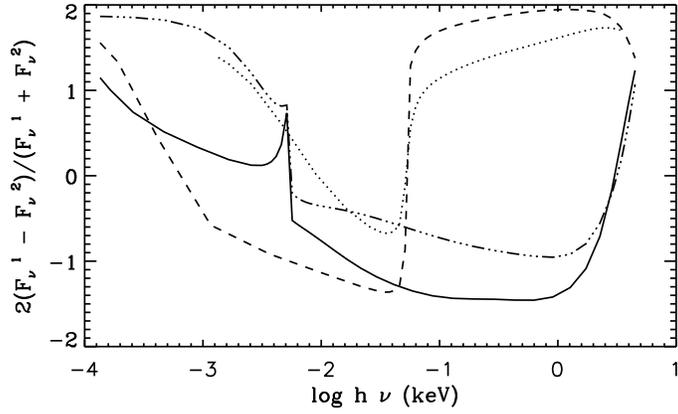,width=10cm}
\caption
{Fraction of polarization
for the same models as in figures \ref{spe12} and \ref{spe13}. Solid
line: model A1; dash--dotted
line: A2; dotted line: A3;
dashed line: A4.
\label{polfig}}
\end{figure}

The fraction of polarization strongly depends on the energy band
and, in the presence of a temperature and density gradient, shows
a variety of different behaviours (see figure \ref{polfig}). Its
sign is determined by the competition between plasma and vacuum
properties in the photospheric layers. However, independently of
the model parameters, the degree of polarization crosses zero at
the very vicinity of the proton cyclotron energy (see eq.
[\ref{criticion}]), where the mode absorption coefficients cross
each other. The bulk of the thermal emission from low--luminosity
accreting NSs falls in the extreme UV/soft X--ray band, which is
subject to strong interstellar absorption, making difficult their
identification. It has been suggested that the detection of the
non--thermal cyclotron emission feature at highest energies may be
a distinguishing signature for most of these low--luminosity
sources (Nelson, \etal \ \cite{nel95:1995}). Our results show that
observations of the proton cyclotron line combined with measures of
polarization may also provide a powerful tool to determine the
magnetic field of the source, even in the absence of pulsations.

\section{Discussion and Conclusions}\label{discuss}

We have discussed the spectral distribution of the radiation
emitted by a static, plane--parallel atmosphere around a strongly
magnetized neutron star which is heated by
accretion. Synthetic spectra have been computed solving the full
transfer problem in a magnetoactive plasma for several values of the
accretion luminosity and of the star magnetic field. In particular, we
explored the low luminosities ($L\sim 10^{30}$--$10^{33}$ erg/s),
typical e.g. of isolated accreting NSs, and found that model spectra show
a distinctive excess at low energies over the blackbody distribution which
best--fits the X--rays. The energy $\hbar\omega_{ex}$ at which the spectrum
rises
depends both on the field strength and the luminosity, but for
$B\approx
10^{12}$--$10^{13}$ G $\hbar\omega_{ex}\sim 10$ eV, so that the optical
emission of an
accreting, NS is enhanced with respect to what is expected
extrapolating the X--ray spectrum to optical wavelengths. The presence of
an optical/UV bump is due to the fact that the low--frequency radiation
decouples at
very low values of the scattering depth in layers where the gas is kept at
larger temperatures by Compton heating.
Since a hot outer zone is exhibited by accreting and not by cooling
atmospheres, the optical excess becomes a
distinctive spectral feature of NSs accreting at low rates.

Despite present results are useful in shedding some light on the emission
properties of accreting, magnetized neutron stars, many
points still need further
clarification before the problem is fully
understood and some of the assumptions on which our investigation was
based deserve a further discussion.
In this paper we have considered only fully
ionized, pure hydrogen atmospheres. In the case of cooling NSs, the
emitted spectra are strongly influenced by
the chemical composition of the surface layers which results
from the supernova explosion and the subsequent envelope
fallback. Present uncertainties motivated several authors to
compute cooling spectra for different abundances (see e.g. Miller,
\& Neuh\"auser \cite{mn91:1991}; Miller \cite{mil92:1992}; Pavlov,
\etal \ \cite{pszm95:1995}; Rajagopal, Romani,
\& Miller \cite{rrm97:1997}) and led to the
suggestion that the comparison between observed and synthetic X--ray
spectra may probe the chemistry of the NS crust (Pavlov, \etal \
\cite{pztn96:1996}).
The assumption of a pure hydrogen atmosphere, although crude, is not
unreasonable for an accretion atmosphere. In this case, in fact,
incoming protons and spallation by energetic particles in the
magnetosphere may enrich the NS surface with light elements (mainly H)
and, owing to the rapid
sedimentation, these elements should dominate the photospheric layers
(Bildsten, Salpeter, \& Wasserman \cite{bsw92:1992}).

Even in this simplified picture, however, the
assumption of complete ionization may not provide an entirely
realistic description.
In a strongly magnetized plasma atomic binding is greatly enhanced, mainly
for light elements. For a typical field of $\sim 10^{12}$ G,
the ionization potentials for the hydrogen ground state are $\sim 100-300$
eV, moving the photoionization thresholds into the soft X--rays
(see e.g. Ruderman \cite{r72:1972}; Shibanov
\etal \ \cite{sh92:1992}). The role of photoionization and of pressure
ionization in a magnetized hydrogen atmosphere has been investigated by a
number of authors (e.g Ventura, Herold, Ruder, \& Geyer \cite{vhrg92:1992};
Potekhin, \& Pavlov \cite{pp97:1997}; Potekhin,
Shibanov, \& Ventura \cite{psv98:1998}). The neutral fraction, $f_H$,
reaches a peak at
densities $\approx 1 \ {\rm g/cm}^3$ then decreases due to
pressure ionization and turns out to be highly
dependent on the temperature. While $f_H$ never
exceeds a few percent at $T = 10^6$ K, it becomes as large as
$\sim 80 \%$ for $T = 10^{5.5}$ K and $B=10^{13}$ G.
Although in the models we presented here such low temperatures are only
reached in relatively low--density layers (see figure \ref{tempfig}),
hydrogen ionization equilibrium should be properly included in a more
detailed analysis.

The effects of different orientations of the magnetic field through the
atmosphere have been also neglected by keeping $\bf{B}\, ||\, \bf{n}$, a
key simplifying assumption which allowed us to solve the transfer problem
in one spatial dimension. Clearly, such
an approximate description is valid only if the size of the
emitting caps is small.
Due to the intrinsic anisotropy of a magnetized medium,
the emerging flux is expected to depend on the angle $\theta_B$ between
the magnetic field and the normal to the surface. Shibanov \etal \
(\cite{sh92:1992}) estimated that in a cooling atmosphere the flux at 1
keV from a surface element perpendicular to the field may exceed that
of an element parallel to ${\bf B}$ by nearly 50\%.
This result shows that a
non--uniform magnetic
field may lead to a significant distortion in the emerging spectra.
Moreover, if the orientation of $B$ varies along the atmosphere,
tangential components of the radiative flux may induce some meridional
circulation to maintain the heat balance (Kaminker, Pavlov, \& Shibanov
\cite{kps82:1982}).

When the finite size of the emitting regions is accounted for, the problem
can not reduced to a plane--parallel geometry and its
solution necessarily demands for multidimensional transfer
algorithms (e.g. Burnard, Klein, \& Arons \cite{bka88:1988},
\cite{bka90:1990}; Hsu, Arons, \& Klein \cite{hak97:1997}).
Calculations performed so far were
aimed to the solution of the
frequency--dependent radiative problem on a fixed background
or of the full
radiation hydrodynamical problem in the frequency--integrated case.
Numerical codes for the solution of the full transfer problem in
axially symmetric media under general conditions are now
available, see e.g. ZEUS (Stone, \& Norman \cite{sn92:1992}),
ALTAIR (Dykema, Castor, \& Klein \cite{dck96:1996})
and RADICAL (Dullemond, \& Turolla \cite{dt99:1999}). Their
application to the transfer of radiation in accretion atmospheres around
magnetized NSs can add new insights on the properties of the emitted
spectra.

\acknowledgments

We are grateful to G.G. Pavlov and to V.E. Zavlin for several very
helpful discussions,
and to A.Y. Potekhin for providing us with the expressions for the ion
opacities. We also
thank S. Rappaport for calling our attention to some useful references.

\appendix
\section{Radiative Processes in a Magnetized Medium}\label{appa}
\subsection{Electron Scattering}

The electron contribution to the scattering source terms has been
evaluated using the expression of the differential cross
section, $d\sigma^{ij}/d \Omega$, discussed by Ventura (\cite{v79:1979})
and
Kaminker, Pavlov, \& Shibanov (\cite{kps82:1982}).
The electron contribution to $K_s^{ij}$
can be simply written as
\be
\label{emi}
K_{s,e}^{ij} =  { 1 \over m_p \kes } \int d \phi'
\dert{\sigma^{ij}}{\Omega } \approx  { 3 \over 4}
\sum_{\alpha=-1}^1
\left | e^{j'}_{\alpha} \right |^2
\left | e^{i}_{\alpha} \right |^2 { 1 \over  \left ( 1 +\alpha u^{1/2}
\right )^2 + \gamma_r^2  } \, .
\ee
where $u = \omega_{c,e}^2/\omega^2$, the $\ e^{i}_{\alpha}$ are
the components of the normal mode
unit polarization vector in a coordinate frame with the
$z$--axis
along $\bf B$ and $\gamma_r=(2/3)(e^2/m_ec^3) \omega$ is the radiation
damping.
Further integration over $\theta'$ gives the total
opacities
\be
k_{s,e}^i = \sum_{j=1}^2 \int K_{s,e}^{ij}\, d\mu'
\approx
\sum_{\alpha=-1}^1
\left | e^{i}_{\alpha} \right |^2 { 1 \over  \left ( 1 +\alpha u^{1/2}
\right )^2 + \gamma_r^2  } \, .
\ee

For energies near the proton cyclotron frequency $\omega_{c,p} =
(m_e/m_p) \omega_{c,e}$, Thomson scattering on ions becomes
important. The corresponding opacity has been derived by Pavlov
\etal \ (\cite{pszm95:1995}) in a relaxation--time approximation
and it is
\be
K_{s,p}^{ij} \approx { 3 \over 4} \mu_m^2
\sum_{\alpha=-1}^1
\left | e^{j'}_{\alpha} \right |^2
\left | e^{i}_{\alpha} \right |^2 { 1 \over  \left ( 1 - \alpha u_p^{1/2}
\right )^2 + \mu_m^2 \gamma_r^2  } \, .
\ee
\be
k_{s,p}^i \approx \mu_m^2 \sum_{\alpha=-1}^1 \left |
e^{i}_{\alpha} \right |^2 { 1 \over  \left ( 1 - \alpha u_p^{1/2}
\right )^2 + \mu_m^2 \gamma_r^2  } \, , \ee where $\mu_m =
m_e/m_p$ and $u_p = \omega_{c,p}^2/\omega^2$. The total opacities
$K^{ij}_s$ and $k_s^i$ appearing in equations (\ref{tr1}) are then
evaluated by adding the contributions of the two species.

\subsection{Bremsstrahlung}\label{brem}

The electron and proton contributions to free--free opacity have a
structure similar to that discussed  for scattering. In a pure
hydrogen plasma, they are given by (see Pavlov, \& Panov
\cite{pp76:1976}; M\'esz\'aros \cite{mes92:1992}; Pavlov \etal \
\cite{pszm95:1995})
\be
k_{ab,e}^i \approx {\kff \over \kes}
\sum_{\alpha=-1}^1
\left | e^{i}_{\alpha} \right |^2 { g_{\alpha} \over  \left ( 1 +\alpha
u^{1/2}
\right )^2 + \gamma_r^2  }
 \, ,
\ee
\be
k_{ab,p}^i \approx {\kff \over \kes} \mu_m^2 \sum_{\alpha=-1}^1
\left | e^{i}_{\alpha} \right |^2 { g_{\alpha} \over  \left ( 1 -
\alpha u^{1/2} \right )^2 + \mu_m^2 \gamma_r^2  } \ee
where
\be
\kff = 4 \pi^2 \alpha_F^3 {\hbar^2 c^2 \over m_e^2} {n_e^2 \over v_T
\omega^3} \left [ 1 - \exp \left ( - \hbar \omega / k T \right )
\right ]  \, ,
\ee
$g_{0} = g_{||}$, $g_{-1} = g_{+1} = g_{\perp}$, and $g_{||}$,
$g_{\perp}$ are the modified Gaunt factors, which account for the
anisotropy induced by the magnetic field. The quantity
$\kff$ is the free--free opacity of a non--magnetic plasma apart
from a factor $(4 \pi /3 \sqrt 3)g$, where $g$ is the unmagnetized Gaunt
factor.
The total absorption opacity can be then evaluated by summing over the
two species.

The modified Gaunt factors were computed evaluating numerically
their integral form as given by Pavlov, \& Panov
(\cite{pp76:1976}). At low temperatures and small frequencies ($u
\gg 1$), where direct numerical quadrature becomes troublesome,
the Gaunt factors have been obtained from the simpler formulas  by
Nagel (\cite{n80:1980}; see also M\'esz\'aros \cite{mes92:1992};
Rajagopal, Romani, \& Miller \cite{rrm97:1997} and references
therein). In order to decrease the computational time, the Gaunt
factors have been evaluated once for all over a sufficiently large
grid of temperatures and frequencies. In the transfer calculation
they are then obtained at the required values of $T$ and $\omega$
by  polynomial interpolation.

\subsection{Vacuum Effects and Mode Switching}\label{vacuum}

The opacities of a real plasma
start to change, due to the vacuum corrections in the
polarization eigenmodes, when the field approaches the  critical value
$B_c=m_e^2c^3/\hbar e \simeq 4.41
\times 10^{13}$ G. The vacuum contribution has ben included modifying
the expressions for the $e_{\alpha}^i$ as discussed by Kaminker, Pavlov,
\& Shibanov (\cite{kps82:1982}), and is controlled by
the vacuum parameter $W$
\be
W = \left({ 3 \times 10^{28} \, {\rm cm^{-3}} \over n_e }\right) \left (
{B \over B_c }\right )^4 \, .
\ee

The inclusion of vacuum and of the protons produces the breakdown
of the NM approximation  near the mode collapse points (MCPs; see
e.g. Pavlov, \& Shibanov \cite{ps79:1979}; M\'esz\'aros
\cite{mes92:1992} and references therein). For $W>4$, or $\rho <
\rho_{vac} = 3.3\times 10^{-3}(B/10^{12}\, {\rm G})^4 \ {\rm
g/cm}^3$, the MCPs appear at the two critical frequencies

\be\label{critic} \omega_{c1,2}^2 = { 1 \over 2}
\omega_{c,e}^2\left[1\pm \left(1-\frac{4}{W}\right)^{1/2}\right]\,
. \ee MCPs play an important role in the transfer of radiation
through a magnetized medium, since  the absorption coefficients of
the two modes either cross each other or have a close approach,
depending on the angle. Following the discussion by Pavlov, \&
Shibanov (\cite{ps79:1979}), under the typical conditions at hand
mode switching is likely to occur at nearly all values of $\mu$.
For this reason and for the sake of simplicity, in the present
calculation we assumed mode switching for any value of $\mu$ at
the two vacuum critical frequencies.

As shown by Bulik, \& Pavlov (\cite{bp96:1996}) in a fully ionized
hydrogen plasma  the presence of protons introduces (even in the absence
of vacuum) a new MCP at
\be\label{criticion}
\omega_{c,3} = \frac{\omega_{c,p}}{\sqrt{1-\omega_{c,p}/\omega_{c,e}+
\omega_{c,p}^2/\omega_{c,e}^2}}\simeq
\omega_{c,p}\left(1+\frac{m_e}{2m_p}\right)\,.
\ee
At $\omega=\omega_{c,3}$ the opacity coefficients cross (Zavlin,
private communication), and the MCP related to the proton contribution is
again a mode switching point. Although we are aware of no detailed
calculation of the
polarization modes in a ``protons + electrons + vacuum'' plasma, it seems
natural to assume that, at least if $\omega_{c,2} > \omega_{c,3}$, the
proton contribution to the ``vacuum'' term
may be safely neglected, being a function of the ratio
$m_e/m_p$.

\section{Stopping Depth and Heating Rate in a Magnetized Atmosphere}
\label{appb}

In the unmagnetized  case, under the assumption that all the
proton stopping power is converted into electromagnetic radiation
within the atmosphere, $W_H$ can be approximated as (Alme, \&
Wilson \cite{aw73:1973}; ZTZT) \be\label{wh} W_H \approx
\cases{\displaystyle { {y_G\kes} \over 8\pi R^2 \tau_s f_A} \left
[L(0) - L(\tau_s) \right ] {{f(x_e)}\over{[1 -
(1-v_{th}^4/v_{ff}^4)(\tau/\tau_s)]^{1/2}}}& $\tau < \tau_s$\cr
&\cr 0 & $\tau \geq \tau_s$\cr} \ee where  $f_A$ is the fraction
of the star surface covered by accretion, $\tau_s$ is the proton
stopping depth and $v_{th}^2/v_{ff}^2 = 3kT(\tau_s) R/(2 m_p G M)$
is the squared ratio of the proton thermal velocity to the
free--fall velocity. For cold atmospheres in which the proton
kinetic energy is much larger than the thermal energy, $f(x_e)$
can be safely taken to be unity for all practical purposes.

Infalling protons loose their energy to electrons via Coulomb
collisions and the generation of collective plasma oscillations
and $\tau_s$ can be approximated as (e.g. Zel'dovich, \& Shakura
\cite{zs69:1969}; Nelson, Salpeter, \& Wasserman
\cite{nel93:1993}) \be\label{taus} \tau_s\approx
\frac{1}{6}\frac{m_p}{m_e}\frac{v_{ff}^4}{c^4\ln\Lambda_c}\simeq
2.6 \, \left(\frac{M}{M_\odot}\right)^2\left(\frac{R}{10^6\, {\rm
cm}}\right)^{-2} \left(\frac{10}{\ln\Lambda_c}\right) \ee where
$\ln\Lambda_c$ is a (constant) Coulomb logarithm.

The proton stopping process in strongly magnetized atmospheres
presents a few differences, and has been discussed in detail by
Nelson, Salpeter, \& Wasserman (\cite{nel93:1993}). Now the proton
stopping depth, $\tau_B$, is larger than $\tau_s$, essentially
because the magnetic field reduces the effective Coulomb logarithm
perpendicular to the field. The expression for the proton stopping
power retains, nevertheless, the same form as in the unmagnetized
case. In the present calculation we used their approximate
expression which relates $\tau_B$ to $\tau_s$ at different field
strengths \be\label{taub} \frac{\tau_B}{\tau_s} \approx
\cases{\displaystyle {\frac{\ln\Lambda_c}{\ln(2n_{max})} }&
$n_{max}\gtrsim 1$\cr &\cr 2\ln(m_p/m_e)\simeq 15 &
$n_{max}\lesssim 1$\cr} \ee $n_{max}=
m_ev_{ff}^2/(2\hbar\omega_{c,e})\simeq$ $ 6.4\,
(M/M_\odot)(R/10^6\, {\rm cm})^{-1} (B/10^{12}\, {\rm G})^{-1}$.
Since proton--proton interactions limit the stopping length to the
mean--free path for nuclear collisions, which corresponds to
$\tau_{pp} \sim 22$ (M\'esz\'aros \cite{mes92:1992}), if the value
of $\tau_B$ which follows form equation (\ref{taub}) exceeds
$\tau_{pp}$, the latter is used instead.

In the very strong field limit ($n_{max}\lesssim 1$), expression
(\ref{wh}) for the heating rate is still valid, provided that
$\tau_s$ is replaced by $\tau_B$. The situation is more
complicated in the moderate field limit ($n_{max}\gtrsim 1$). In
this case, not only $\tau_s$ must be replaced by $\tau_B$, but a
further effect must be accounted for. Now a sizeable fraction,
$\sim 1 - 1/\ln(2n_{max}$), of the initial proton energy goes into
excitations of electrons Landau levels. The cyclotron photons
produced by radiative deexcitation will be partly thermalized by
absorption and Compton scattering while the remaining ones escape
forming a broad cyclotron emission feature (Nelson, \etal \
\cite{nel95:1995}). To account for this we take in the moderate
field limit, \be\label{wmf} W_{MF} = (1-f)W_H +
fW_H(\Delta\epsilon/\epsilon) \ee where $f$ is the fraction of the
initial proton energy which goes into Landau excitations and
\be\label{fracgain} \frac{\Delta\epsilon}{\epsilon}\approx
\frac{(\hbar\omega_{c,e}/m_ec^2)\tau^2}{1+(\hbar\omega_{c,e}/m_ec^2)\tau^2}
\ee is the electron fractional energy gain due to repeated
scatterings. Strictly speaking, equation (\ref{fracgain}) is valid
only for photons produced below $\tau_C\sim
(\hbar\omega_{c,e}/m_ec^2)^{-1/2}\simeq 7\, (B/10^{12}\, {\rm
G})^{-1/2}$, because only these photons loose a significant
fraction of their energy in Compton collisions with electrons. We
have also to take into account that cyclotron photons produced at
depths greater than the thermalization depth, $\tau_{th}\sim 16\,
(B/10^{12}\, {\rm G})^{7/6}(M/M_\odot)^{1/3}(R/10^{6}\, {\rm
cm})^{2/3} (T/10^{7}\, {\rm K})^{1/3}$, are absorbed. In our
models, however, $\tau_B$ is always less than $\tau_{th}$, so we
do not need to worry about free--free absorption in (\ref{wmf}). A
parabolic fit to the sum of first two curves ($0\to 1$ and $0\to
2$ Landau transitions) in figure 4 of Nelson, Salpeter, \&
Wasserman (\cite{nel93:1993}) is very accurate and gives
\be\label{ffit} f(\tau ) = 6.1\times 10^{-4} + 6.5\times
10^{-2}(\tau/\tau_s) + 8.7\times 10^{-3}(\tau/\tau_s)^2\, . \ee

\clearpage

\begin{deluxetable}{cccccc}
\tablecolumns{6}
\tablewidth{0pc}
\tablecaption{Model Parameters\label{table1}}
\tablenum{1}
\tablehead{
\colhead{Model}&
\colhead{$B$} &
\colhead{$L_\infty$} &
\colhead{$L(\tau_B)/L(0)$} &
\colhead{$\dot M$} &
\colhead{$f_A$} \\
\colhead{} &
\colhead{$10^{12}$ G} &
\colhead{$10^{33}$ erg s$^{-1}$} &
\colhead{} &
\colhead{$10^{12}$ g s$^{-1}$} &
\colhead{}
}
\startdata
 A1 & 1 & 0.1 & 0.17 & 0.75 & 0.01\\
 A2 & 1 & 3.7 & 0.25 & 25 & 0.01\\
 A3 & 10 & 0.2 & 0.28 & 0.13 & 0.01\\
 A4 & 10 & 6.3 &  0.33 & 38 & 0.01\\
 A5 & 0 & 4.3 & 0.29 & 28 & 0.01\\
 C1 & 1 & 0.2 & 1 & 0 & 1\\
 C2 & 9 & 0.52 & 1 & 0 & 1\\

\tablecomments{The first letter in the model identifier refers to
accretion (``A'') or cooling (``C'') atmospheres.}
\enddata
\end{deluxetable}

\clearpage

\begin{deluxetable}{ccccc}
\tablecolumns{4}
\tablewidth{0pc}
\tablecaption{Predicted Excess \label{table2}}
\tablenum{2}
\tablehead{
\colhead{Model}&
\colhead{$\displaystyle{F_{6060}}\over \displaystyle{F^{bb}_{6060}}$} &
\colhead{$\displaystyle{F_{3000}}\over \displaystyle{F^{bb}_{3000}}$} &
\colhead{$T^{bb}_{fit}$ (keV)} \\
}
\startdata
 A1 & 1.60 & 1.36 & 0.25 \\
 A2 & 3.05 & 1.44 & 0.43 \\
 A3 & 1.12 & 1.06 & 0.25 \\
 A4 & 3.44 & 2.21 & 0.57 \\
 A5 & 3.49 & 2.16 & 0.52 \\

\enddata
\end{deluxetable}

\clearpage

\end{document}